\documentclass{article}

\usepackage{arxiv}

\usepackage[utf8]{inputenc} % allow utf-8 input
\usepackage[T1]{fontenc}    % use 8-bit T1 fonts
\usepackage{hyperref}       % hyperlinks
\usepackage{url}            % simple URL typesetting
\usepackage{booktabs}       % professional-quality tables
\usepackage{amsfonts}       % blackboard math symbols
\usepackage{nicefrac}       % compact symbols for 1/2, etc.
\usepackage{microtype}      % microtypography
\usepackage{lipsum}		% Can be removed after putting your text content
\usepackage{graphicx}

\usepackage{wrapfig}

\title{Artificial neural networks for disease trajectory prediction in the context of sepsis}

\date{July 28, 2020}	% Here you can change the date presented in the paper title
\author{ \href{https://orcid.org/0000-0001-8591-9072}{\includegraphics[scale=0.06]{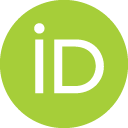}\hspace{1mm}Dale Larie}\thanks{Use footnote for providing further
		information about author (webpage, alternative
		address)---\emph{not} for acknowledging funding agencies.} \\
	Department of Surgery\\
	University of Vermont\\
	Burlington, VT \\
	\texttt{dale.larie@uvm.edu} \\
	\And
	\href{https://orcid.org/0000-0003-4549-9004}{\includegraphics[scale=0.06]{orcid.png}\hspace{1mm}Gary An} \\
	Department of Surgery\\
	University of Vermont\\
	Burlington, VT \\
	\texttt{gary.an@med.uvm.edu} \\
		\And
	\href{https://orcid.org/0000-0003-3224-7617}{\includegraphics[scale=0.06]{orcid.png}\hspace{1mm}Chase Cockrell} \\
	Department of Surgery\\
	University of Vermont\\
	Burlington, VT \\
	\texttt{robert.cockrell@uvm.edu} \\
}

\hypersetup{
pdftitle={A template for the arxiv style},
pdfsubject={q-bio.NC, q-bio.QM},
pdfauthor={David S.~Hippocampus, Elias D.~Striatum},
pdfkeywords={First keyword, Second keyword, More},
}

\begin{document}
\maketitle

\begin{abstract}
{\bf{Introduction:}} The disease trajectory for clinical sepsis, in terms of temporal cytokine and phenotypic dynamics, can be interpreted as a random dynamical system.  The ability to make accurate predictions about patient state from clinical measurements has eluded the biomedical community, primarily due to the paucity of relevant and high-resolution data. 
{\bf{Methods:}} We have utilized two distinct neural network architectures, Long Short-Term Memory and Multi-Layer Perceptron, to take a time sequence of five measurements of eleven simulated serum cytokine concentrations as input and to return both the future cytokine trajectories as well as an aggregate metric representing the patient’s state of health.  We performed this work with two distinct training sets: one consisted of cytokine trajectories starting immediately after the simulated injury and one consisted of cytokine trajectories beginning 24 hours after the simulated injury. 
{\bf{Results:}} The neural networks converged within 50 epochs for cytokine trajectory predictions and health-metric regressions, with the expected amount of error (due to stochasticity in the simulation). The mapping from a specific cytokine profile to a state-of-health is not unique, and increased levels of inflammation result in less accurate predictions and increased amounts of stochasticity. The Multi-Layer Perceptron neural network, trained on trajectories beginning 24 hours post-injury performed the best.  Due to the propagation of machine learning error combined with computational model stochasticity over time, the network should be re-grounded in reality daily as predictions can diverge from the true model trajectory as the system evolves towards a probabilistic basin of attraction. 
{\bf{Discussion:}} This work serves as a proof-of-concept for the use of artificial neural networks to predict disease progression in sepsis.  This work is not intended to replace a trained clinician, rather the goal is to augment their intuition with quantifiable statistical information to help them make the best decisions. We note that this relies on a valid computational model of the system in question, in this case, the innate immune system, as there is not now, nor will there be in the near future, sufficient data to inform a machine-learning trained, artificially intelligent, controller.
\end{abstract}

% keywords can be removed

\section{Introduction}
There are approximately 1 million cases of sepsis in the United States each year, with a mortality rate between 28-50\% \cite{RN1}.  Sepsis is a highly dynamic process, characterized phenotypically by features such as multi-system organ failure, and molecularly, by dysregulation of the body’s internal cytokine signaling network \cite{RN2,RN3,RN4}. While care process improvements in the treatment of sepsis, such as the development of treatment bundles and practice guidelines, have improved clinical outcomes in the past few decades, the search for new drugs to treat the biological-basis of sepsis has been marked by complete failure: there is currently not a single drug approved by the U.S. Food and Drug Administration that targets the underlying pathophysiology of sepsis \cite{RN5,RN6}. One of the major challenges in designing therapies for sepsis is an inability to effectively forecast the disease trajectories of individual patients, thereby limiting the effective sub-stratification of this heterogeneous population into those biologically similar enough to control.  Existing means of classifying sepsis patients, such as with the Sequential Organ Failure Score (SOFA) \cite{RN7} or various biomarker panels \cite{RN8,RN9,RN10}, while potentially useful for coarse-grained outcome risk stratification, are only able to provide population-level projections that cannot effectively be updated to an individual patient’s disease course.  Adding to the limitations of data-centric population-based scoring systems is the inherent stochasticity of the biological processes driving sepsis. The presence of stochasticity in the system governing inflammation makes accurately predicting the entire trajectory of the disease, or accurately predicting the patient state 30 days into the future, given one point of assessment, an impossibility.  

Ultimately, the biological heterogeneity seen clinically is a combination of inter-patient (genetic variability) and intra-patient (stochastic processes) heterogeneity.  The result is that it is impossible to comprehensively enumerate all possible biomarker states and configurations (i.e. phenotypes) that can be generated from a specific systemic perturbation or injury.  The challenge (and solution) is similar to that faced by Q-Learning \cite{RN11} (now Deep Reinforcement Learning); Q-learning is a type of reinforcement learning in which agents determine what action to take (a) by looking up their current state (s) in the lookup table, Q(s,a) that lists the probability of a desirable outcome based on that decision.  Because the lookup table needs to provide this probability to guide the decision process it requires a finite (and computationally tractable) state space. In order to work effectively in continuous (infinite states) search spaces, Q-learning takes utilizes the Universal Approximation Theorem \cite{RN12}, which states that a feed-forward neural network can approximate, to arbitrary fidelity, a real and continuous function. In the case of Q-learning, it is the lookup table that is being approximated; we note that the lookup table does not necessarily meet the strict mathematical definition for continuity, however the technique works in practice. Personalized medicine faces a similar challenge: it is impossible to comprehensively enumerate the set of all possible clinical observables (cytokine profiles, vital signs, etc.) which a patient can present – these state predictions must then be approximated.  Furthermore, being able to eventually discover and evaluate potential therapeutic regimens/control strategies requires the ability to evaluate counterfactuals: e.g. what would have happened had an intervention not been done? The ability to represent counterfactuals requires being able to depict some future horizon of system behavior while accounting for the inherent stochasticity and behavioral heterogeneity of the system.

Due to the stochastic nature of biomedical systems, predictions should be thought of as analogous to predicting the path of a hurricane – making short-term predictions is possible, making long-term (and accurate/precise) predictions is not;  in order to effectively predict the path of a hurricane, and mitigate the risks associated with it, the model is continuously updated and new putative trajectories are calculated. Such an approach requires the ability to capture the dynamic processes that drive the behavior of the system, and as such mandate the use of mechanistic/quasi-mechanistic models that can dynamically generate system trajectories. 

In previous work, we have demonstrated that the cytokine signaling network which controls the inflammatory process can be modeled as a random dynamical system \cite{RN2,RN4}, which is a system that evolves in time according to fixed rules, but also incorporates stochasticity \cite{RN13,RN14}. Knowledge of the underlying cellular and molecular processes of acute inflammation has been used to create a dynamic model, the Innate Immune Response Agent-based Model (IIRABM) \cite{RN15}, that can serve as a proxy model for the development of more advanced prediction and forecasting methods. The IIRABM  has been used to demonstrate the use of in silico clinical trials as a means of evaluating the plausibility of planned potential interventions \cite{RN15}, provided fundamental insights into the mathematical and dynamic properties of sepsis that account for patient heterogeneity \cite{RN2}, demonstrating the futility of standard biomarker-based outcome prediction \cite{RN2}, and served as a proxy model \cite{RN16} for control discovery for sepsis. This most recent control discovery work has employed advanced computational methods such as genetic algorithms/evolutionary computing \cite{RN4} and deep reinforcement learning/artificial intelligence \cite{RN17} to describe what would be required for multi-modal treatment of sepsis. While the IIRABM is nearly 20 years old its central component structure remains valid and has predicted a series of behaviors associated with sepsis that have since been recognized in the subsequent years, specifically the temporal concurrence of pro- and anti-inflammatory cytokine responses (as opposed to sequential pro- and compensatory responses) \cite{RN18,RN19} and the importance of the immunoparalyzed recovery phase of sepsis, particularly with respect to its prolonged duration \cite{RN20,RN21,RN22,RN23}.  The current work utilizes the IIRABM as a means of developing a method for dynamic trajectory prediction through the training and use of artificial neural networks (ANN). 

In the hospital setting, there is a paucity of population-level comprehensive data which can characterize a patient’s health trajectory, causing machine-learning models that are informed by data alone, without an underlying mechanistic model or biological theory, to be brittle and of limited utility \cite{RN24,RN25}.  The use of a properly validated computational model, which can generate unlimited amounts of data, can address this shortcoming, however, in order for multiscale modeling and simulation to be deployed in clinical practice, it must be practical to utilize the models in a clinical setting.  As we have shown in previous work \cite{RN2}, this requires an immense amount of computational power as the simulation has to be repeated for many stochastic replicates.  Compressing/approximating the information and dynamics contained within the computational model using an ANN allows for a computationally cheap and tractable method of rapidly updating predictions about patient disease trajectory as new information becomes available.

\section{Materials and Methods}
\label{sec:headings}

The foresting procedure is divided into two principal tasks: 1) predict future cytokine trajectories in an 11-dimensional space; and 2) regress the overall ‘health’ of the simulation as a function of its current cytokine profile.  Training and validation data was generated using a previously validated computational model of the human immune response to injury, the Innate Immune Response Agent-Based Model (IIRABM) \cite{RN2}.  The training/validation set was composed of cytokine measurements for 11 unique cytokines over 10,000 time-steps in 66,000 in silico patients.  Networks were constructed Using Keras \cite{RN26}, a TensorFlow based deep learning library for Python.

\subsection{Trajectory Forecasting}
In order to forecast future values in the cytokine time series, we utilized long short-term memory (LSTM) recursive neural networks (RNN).  RNNs are different from standard multi-layer-perceptron networks because they have a neural network contained within a cell which takes information from the current input to help determine the adjusted state of the cell based on its current cell state. This adjusted cell state becomes the new cell state, and an output is determined for the network. 

LSTM networks’ memory cells have a unique structure, characterized by an input gate, two update layers, and an output gate to determine the adjusted cell state \cite{RN27}. The memory cells in LSTM networks allow for more long term memory than typical RNNs which make them well suited for time-series analysis and prediction \cite{RN28}. Noting this, an LSTM network will likely be able to predict future cytokine levels, given that they are continuous and previous cytokine levels will likely have a large impact on near-future values.

We constructed a unique network for each cytokine that was to be predicted; each LSTM network takes 5 sequential 11-d cytokine profiles as input and predicts the subsequent value(s). The first three layers of the network are 100-node LSTM layers; the output from these layers are fed into two fully connected layers of 300 and 200 nodes respectively, then to a single output node, resulting in 296,301 trainable parameters. Training data was arranged into 5 sequential 11-dimensional points as training input features and the next 11-dimensional point as the training label. The data was then shuffled to avoid biasing the training. After data preprocessing, 8,576,100 data sequences and labels were used to train the network. The loss metric used to train this network is mean absolute error (MAE), and the Adam optimizer \cite{RN29}. Each network was trained until loss converged to a minimum

For the ultimate utilization of this network, 11 cytokine values are observed for 5 time steps, then a prediction for each of the next values is made using its own LSTM network. This set of 11 observations is combined into one 11-dimensional point, which is then added to the original 5 samples as the next sample. Predictions are made recursively in this manner for 100 time steps after the initial observation. Accuracy of this algorithm was measured using the average MSE across the 11 cytokine values at 1, 2, 3, 4, 5, 10, 25, 50, and 100 time steps after the initial observation.  Prediction variance and error bars were calculated through stochastic variations to the dropout layer \cite{RN30}, as demonstrated with regards to Active Learning for regression in \cite{RN31}.
\subsection{Health Metric Regression	}
The IIRABM uses the ‘Oxygen Deficit’ metric as a measure of health, where a low oxygen deficit is good, and a high oxygen deficit is bad. We note that, both in silico and in vivo, cytokine profiles provide a non-unique mapping to state-of-health (a concept which is more nebulously defined in vivo than in our in silico model). As such, error is expected when attempting to regress from an 11d cytokine profile to a single health metric.
\begin{figure}[h]
\centering
\includegraphics[width=\textwidth]{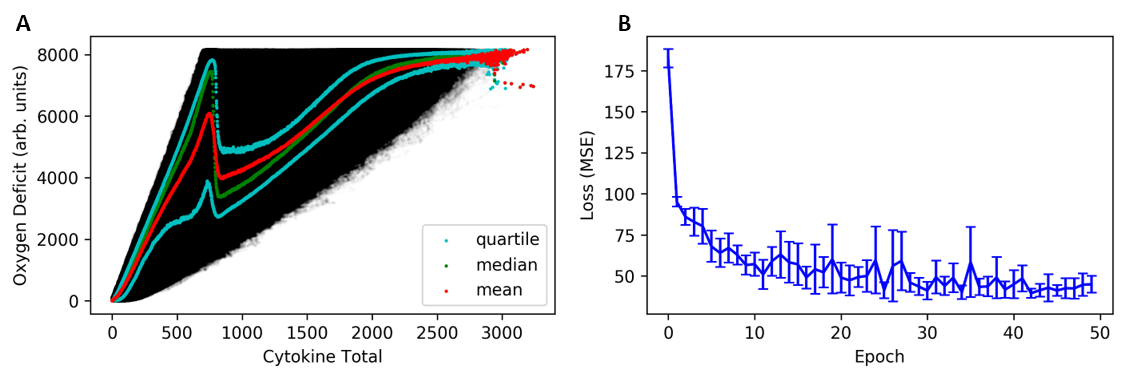}
\caption{
In Panel A we present the variance in oxygen deficit as a function of the sum of cytokine concentrations in the whole area of simulated tissue.  In Panel B, we show the mean absolute error in the regression of the oxygen deficit as a function of cytokine profile, as a function of training epoch.}
\label{fig1}
\end{figure}
In order to perform this regression, we utilized a fully connected deep network which takes an 11-dimensional cytokine vector as input, feeding into two fully connected layers with 1,500 nodes each, then into a layer with 150 nodes, and finally to a single output node. The loss metric used to train this algorithm is MSE. Using the regression network, a prediction of oxygen deficit trajectory can be made from the 11-dimensional matrix created by the LSTM network. Overall accuracy was measured by comparing the oxygen deficit path to the predicted path and calculating the MSE.

The prediction of whether an in silico patient will live or die, or the decision on whether or not pharmacologic therapeutics are more likely to be beneficial than detrimental, is ultimately based on the temporal trajectory of the patient’s state of health (in this case, a measure of systemic oxygen deficit). The predicted trajectory in 11d cytokine space is then fed into the health-metric regression network to forecast the most-likely outcome, time-to-outcome, and time-horizon for potentially effective therapeutic interventions.

Additionally, we created a multi-layer perceptron (MLP) to predict the future oxygen deficit trajectory as a function of past values only, effectively treating the simulation output as a Markov chain.  This network expects an input of 5 sequential oxygen deficit values and will return a single future oxygen deficit value predicted for the next time step. The structure of this MLP begins with a fully connected layer of 1000 nodes, followed by a 1\% permanent dropout layer, then two fully connected layers of 150 nodes each, followed by a single output node. This network was trained using a loss function to minimize MSE. Trajectory prediction for this network is made in the same recursive manner as the cytokine prediction networks.

Lastly as a comparison of the efficacy of LSTM neural networks, MLP prediction networks for each cytokine were also created. Each network functionally acts the same as the LSTM networks, accepting 5 sequential 11-dimensional points in cytokine space and predicting the future value for a single cytokine. Each network has a structure beginning with a fully connected layer of 1000 nodes, followed by a function to flatten the output shape from a 5 by 1000 array to a single vector of length 5000. Next is another fully connected layer of 1000 nodes, then a 1\% permanent dropout layer, feeding into a fully connected layer of 500 nodes, then another fully connected layer of 500 nodes, then finally to a single output node. These networks were trained using a loss function to minimize MSE.

\section{Results}

It is important to note that the map which translates a cytokine profile into its associated oxygen deficit (and vice-versa) is non-unique, and therefore some amount of error is expected and unavoidable. In Figure 1a, we present the variance in oxygen deficit as a function of the sum of cytokine concentrations in the whole area of simulated tissue.  The sum of cytokine concentrations is a coarse metric that roughly represents the amount of inflammation (no distinction is made between pro- and anti-inflammatory signals) and inflammatory signaling present in the model.  This is analogous to what is seen clinically – patients that see ostensibly identical insults/infections/injuries will invariably present a range of responses, in terms of temporal cytokine profiles or other clinical observables (heart-rate, blood pressure, temperature, etc.).

This figure also illustrates a key difference between the structure of the noise in the IIRABM and the stochastic structure in a stochastic differential equation; the noise present in the IIRABM varies spatio-temporally and cannot be represented with a closed-form analytical expression.  Very generally speaking, the reason for this is that when cell-signaling is high, there is lots of activity in the model, and therefore lots of opportunities for stochastic events.  This can be illustrated with a simple thought experiment: consider two system states, one with a single infected cell and one with 10 infected cells, and each infected cell has a probability of infecting a single neighbor, and some probability of healing.  If we evolve the simulation a single time step, system 1 can have 0, 1, or 2 infected cells, while the range of infected cells in system 2 can vary from 0 to 20 (depending on the spatial configuration of the infected cells).  In Figure 1b, we show the mean absolute error as a function of training epoch when training the health regression neural net.  The error quickly converges to a minimum with a relatively constant error of approximately 200 units (on a scale of 8160), with the caveat that the predicted error would be lower when the true oxygen deficit is lower, and higher when the true oxygen deficit is higher.
\begin{wrapfigure}{r}{0.5\textwidth}
\includegraphics[width=0.5\textwidth]{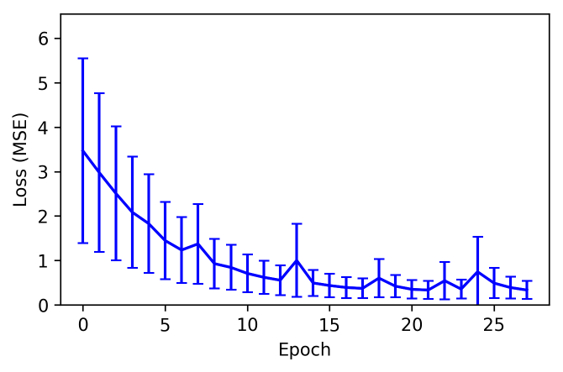}
\caption{ The mean squared error (in arbitrary units) as a function of training epoch for TNF$\alpha$.  The training of this cytokine prediction network was representative of all simulated cytokine prediction networks.}
\label{fig2}
\end{wrapfigure}

\begin{wrapfigure}{l}{0.5\textwidth}
\includegraphics[width=0.5\textwidth]{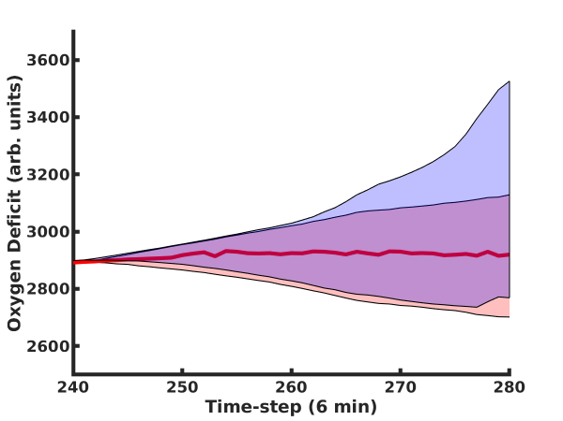}
\caption{
The future health-trajectory probability cloud predicted using the MLP network is shaded in red; the future health-trajectory probability cloud predicted using the LSTM network is shaded in blue; the true trajectory is plotted in red.
}
\label{fig3}
\end{wrapfigure}

Cytokine trajectories present similar stochastic properties as the oxygen deficit: when levels are high, the plausible range of cytokine expression for the subsequent time step is larger than when levels are low.  We present the mean squared error (in arbitrary units) as a function of training epoch for TNF$\alpha$, which is representative of the full cytokine set in Figure 2.  Once again, the network quickly converges to a low and constant level of error.  We note that the total error quickly and significantly increases as we extend the time-prediction horizon past 100 time-steps.  This distinguishes this methodology from that of ML-augmented surrogate modeling \cite{RN32} because we do not claim the ability to accurately represent the entire course of a sepsis disease trajectory (up to 90 days in our computational model) using neural-network approximations.

The use of the dropout layer allows for the simple creation of an ensemble of predictive networks by stochastically varying the specific node(s) in the layer that are dropped out.  Using this, we have visualized the probability cloud for future health trajectories generated using the MLP network (shaded in red), future health trajectories generated using the LSTM network (shaded in blue) and plotted the true trajectory (red line) in Figure 3.  This figure visualizes a single prediction iteration (predict future cytokine trajectory, regress state of health) for the above-described workflow.  As new data is fed into the model about the true trajectory of the system, the forecast cloud is updated.  The actual health trajectory typically lies in the center of the probability cloud, which is a clear benefit of the ensemble approach.  In Figure 4, we display the probability cone for the entire simulation run, starting at the 240th time step, and then updating the trajectory cone on every subsequent time step.

In Figure 5, Panel A, we contrast predictions that considered the full time evolution of the system when training the neural network model, shaded in red, with predictions that only used training data collected after the 240th time step, representing approximately 1 day.  The network that only utilizes data collected more than 24 hrs post-injury performs substantially better.  This is primarily due to the massive amount of stochasticity introduced at the time of injury; the degree of this stochasticity is significantly larger in magnitude than later in the simulation, as discussed below. In Panel B, we display the same oxygen-deficit probability cone as in Panel A, however also reseed the simulation’s random number generator at this time step to generate 100 stochastic replicates of the time evolution of that specific instantiation of the IIRABM.  We see that the predicted probability cone has a greater spread than the actual probability cone, however we note that the MLP predictor is constantly updating its trajectory predictions: the set of observations $\{t_{-5}^a,t_{-4}^a,t_{-3}^a,t_{-2}^a,t_{-1}^a\}$, where $t_{-5}^a$ has the superscript, ‘a’, representing an actual observation, and the subscript, ‘-5’ to denote that the time point is 5 points prior to the starting reference point, is used to predict $t_0^p$, where the superscript, ‘p’, indicates a predicted observation.  Eventually, the set of points used to generate the prediction will consist entirely of previously predicted points, allowing for the compounding of any errors.  In Panel C, we show the same probability cone as in Panel A, however this time, we have reseeded the simulation every 100 time steps at t=1100 to t=1800, for 100 stochastic replicates each.  This is a more direct comparison since the MLP predictor effectively reseeds itself every time step.  We observe that the actual probability cone is significantly wider than in Panel B, but still not as wide as the predicted cone.  This is discussed in detail below.
\begin{wrapfigure}{r}{0.5\textwidth}
\includegraphics[width=0.5\textwidth]{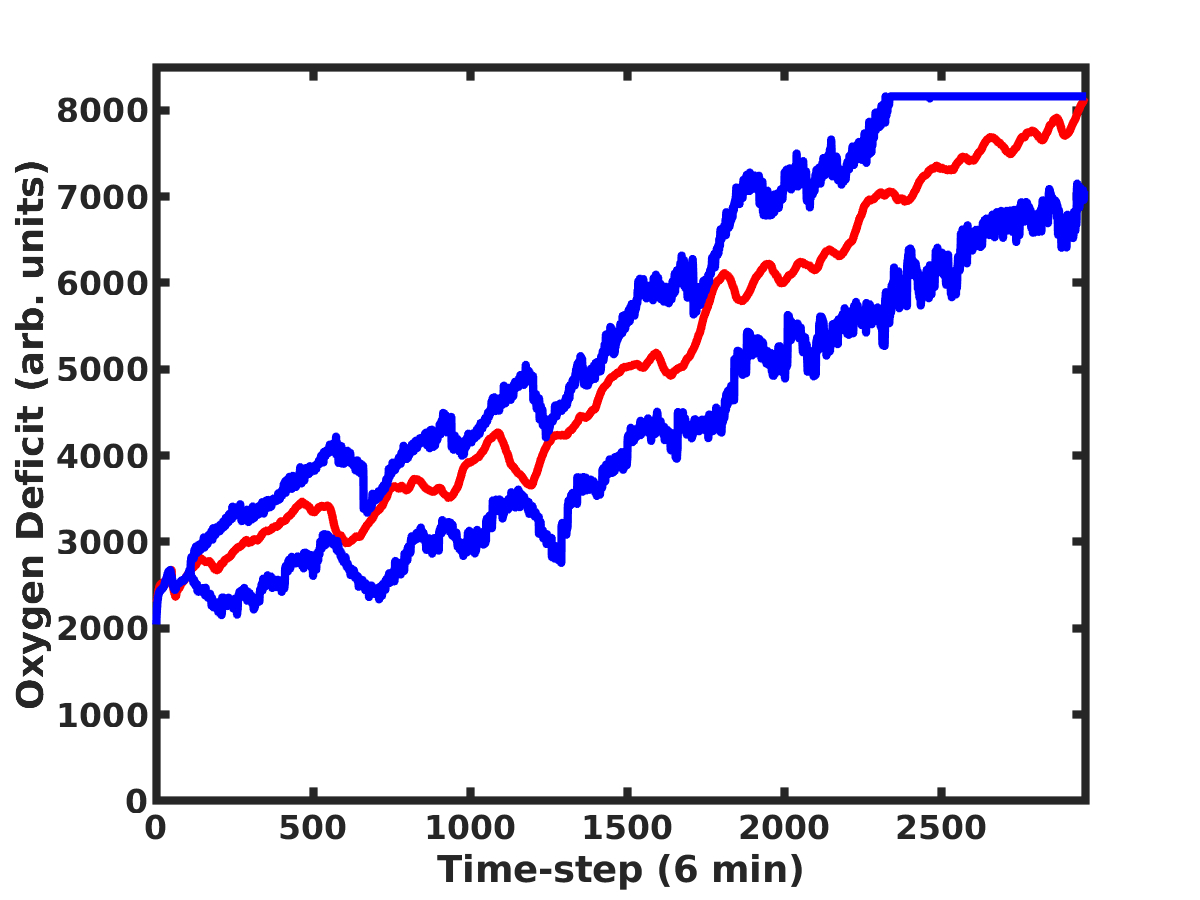}
\caption{
The upper and lower boundaries for the future health-trajectory probability cone are indicated by the blue lines; the actual trajectory is plotted with a red line.  Predictions began at t=200 and were update upon every time-step.
}
\label{fig4}
\end{wrapfigure}

\section{Discussion}
The MLP predictor, treating the oxygen deficit as a Markov chain, performs better than using an LSTM to predict future state-of-health trajectories, however this does not represent a failure of the LSTM method (or indicate superiority over the MLP).  This is best illustrated in figure 1, which illustrates the non-unique mapping between a specific cytokine profile and a state of health, which is well-known clinically \cite{RN33}.  The accuracy of the cytokine trajectory predictions, shown in Fig. 2, is high, but even with an accurate prediction of the future cytokine profile, turning that profile into an informative state of health prediction is not possible.

Additionally, we note that the predictor performs better when using training data starting 1 day after the simulated injury perturbs the system, and the reasoning for this is similar to that above, in that the cytokine and histological dynamics are dominated by stochasticity. When the simulated injury occurs, a large, contiguous, area of tissue is injured with a homogenous injury, representing a significant perturbation to the system; thus, early simulation behaviors contain a significant amount of stochasticity, leading the training data to be less informative as to the true mechanisms which underlie the dynamics of the simulation.   

Recognizing that there are configurations in which the system is more or less strongly influenced by randomness can also help to explain why treating the global simulation output as a Markov chain (or full cytokine trajectory output as a Markov random field, as we have described in \cite{RN2}) is only an approximation.  The full simulation, which takes place on a two-dimensional grid, is memoryless, and begins with a homogenous injury.  However, as the injury evolves, the spatial distribution of damage or of various cytokine concentrations begins to vary, due to both stochastic and deterministic influences.  All of this information about spatial heterogeneity is lost when it is collapsed into a single trajectory.  A Markov Transition Matrix (or kernel) could be constructed for the simulation output that would be true in the comprehensive sense, e.g., when considering all possible trajectories and model configurations, however the utility of this information becomes more limited the farther out the prediction goes, as seen in Fig 5.  
\begin{figure}[h]
\centering
\includegraphics[width=\textwidth]{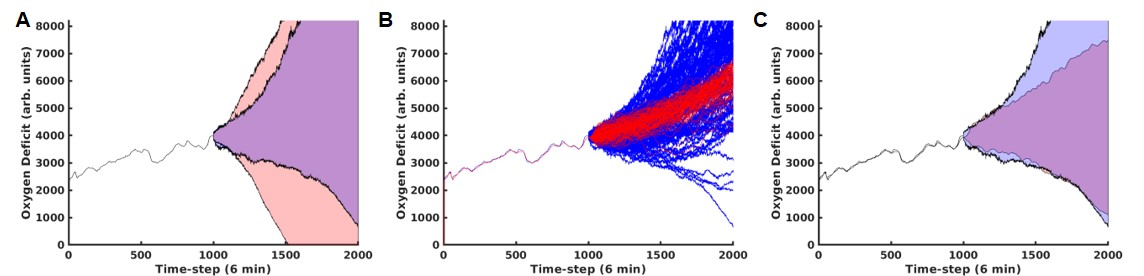}
\caption{
In Panel A, we contrast the future health-trajectory probability cones with networks that used data collected in the first 24 hours (shaded in red) and networks that excluded the first-day data (shaded in blue); we note that the blue area appears purple as it is entirely contained within the red area.  In Panel B, we present 100 stochastic replicates of the actual simulation health-trajectory, reseeded at the time of the first prediction (red) compared with the predicted probability cone trajectories (blue).  In Panel C, we generated the simulation trajectory cone through reseeding the simulation's random number generator at the upper and lower boundaries of the trajectory cone every 100 time steps from t=1100 to t=1800.
}
\label{fig5}
\end{figure}
The model reseedings in Figure 5 indicate to use that the model is in a very deterministic configuration. Due to the spatial distribution of the injury, there is essentially no chance that it will heal the in-silico patient back to full health, while also being in no danger of an immediate/near-term death.  Essentially, the simulations finds itself entirely under the influence of a single Probabilistic Basin of Attraction; this is discussed in detail in \cite{RN2}.  So while the fully spatially realized simulation does not have a memory (and can safely be treated as a Markov process), the historical paths of the aggregate cytokine/health trajectories are important.

The failure to identify effective drugs to treat sepsis is ultimately a failure to account for the heterogeneity of the state space for sepsis and the non-uniqueness of mapping from state space to trajectory space: without understanding the potential future histories of an individual patient from any point in time there can be no rationally justified attempt at controlling or steering that patient’s eventual outcome. We have proposed that mechanism-based multi-scale computational models (as defined by the National Institutes of Health Interagency Modeling and Analysis Group https://www.imagwiki.nibib.nih.gov/content/multiscale-modeling-msm-consortium) can serve as proxy systems that can address the “Denominator Problem” that arises out of the non-uniqueness of the mapping between system state and behavior and the inevitable sparsity of biological data \cite{RN34}; we pose that the IIRABM represents one example of a proxy model for sepsis. The ability to predict requires an additional layer of surrogate models in order to render such prediction clinically tractable, and the complexity of the dynamic structure of inflammation/sepsis calls for the use of ANNs for this purpose. While there have been some attempts to use ANNs to serve as surrogates for multiscale models [35], these do not directly address complex agent-based models, such as the IIRABM, which are explicitly used because they have no equivalent equation-based formulation. The key aspect of the current work is that it represents a first-step of a development workflow that integrates mechanism-based simulation with machine learning in order to train predictive ANNs that can inform what sort of sensing technology and capabilities need to be developed in the real world. Rather that sporadically connected (at best) suggestions of which mediators to target, and at what time interval, the implementation of workflows using model-driven investigation will bring biomedicine in line with nearly every other technological and scientific field.

\section{Acknowledgments}
This work was supported by National Institutes of Health grant 1RO1GM115839-01.  Additionally, this research used high performance computing resources provided by the Vermont Advanced Computing Core (VACC). 

\bibliographystyle{plos2015}
\bibliography{SepsisTrajectory}

\end{document}